\documentclass[twocolumn,showpacs,preprintnumbers,amsmath,amssymb,pra]{revtex4}

\usepackage{graphicx}
\usepackage{dcolumn}
\usepackage{bm}
\usepackage{mathrsfs}
\usepackage{amssymb}
\usepackage{amsmath}
\newcommand{\ave}[1]{\langle #1 \rangle}
\newcommand{\bra}[1]{\langle #1|}
\newcommand{\ket}[1]{| #1 \rangle }
\newcommand{\ip}[2]{{\langle #1|}{ #2 \rangle }}
\newcommand{\tr}[1]{{\rm tr}[#1]}
\newcommand{\be}{\begin{eqnarray}}
\newcommand{\ee}{\end{eqnarray}}

\newcommand{\E}{{\sf E}}

\newcommand{\cS}{{\cal S}}
\newcommand{\cH}{{\cal H}}

\newcommand{\cO}{{\cal O}}

\begin{document}

\title{Probability-based comparison of quantum states}

\author{Sergey N. Filippov$^1$ and M\'{a}rio Ziman$^{2,3}$}
\affiliation{
$^1$Moscow Institute of Physics and Technology, Moscow Region, Russia \\
$^2$Institute for Theoretical Physics, ETH Zurich, 8093 Zurich, Switzerland\\
$^3$Institute of Physics, Slovak Academy of
Sciences, Bratislava, Slovakia }

%\date{February 5, 2012}

\begin{abstract}
We address the following state comparison problem: is it possible
to design an experiment enabling us to unambiguously decide (based
on the observed outcome statistics) on the sameness or difference
of two unknown state preparations without revealing complete
information about the states? We find that the claim ``the same"
can never be concluded without any doubts unless the information
is complete. Moreover, we prove that a universal comparison (that
perfectly distinguishes all states) also requires complete
information about the states. Nevertheless, for some measurements,
the probability distribution of outcomes still allows one to make
an unambiguous conclusion regarding the difference between the
states even in the case of incomplete information. We analyze an
efficiency of such a comparison of qudit states when it is based
on the SWAP-measurement. For qubit states, we consider in detail
the performance of special families of two-valued measurements
enabling us to successfully compare at most half of the pairs of
states. Finally, we introduce almost universal comparison
measurements which can distinguish almost all non-identical states
(up to a set of measure zero). The explicit form of such
measurements with two and more outcomes is found in any dimension.
\end{abstract}


\maketitle

%-------------------------------------------------------------------------------------------

\section{\label{section:introduction} Introduction}
The exponential scaling of the number of parameters describing
multipartite quantum systems stands behind the potential power of
quantum information processing. However, the same feature makes a
complete characterization (tomography) of unknown quantum devices
intractable. Therefore, it is of practical interest to understand
which properties of physical systems require the full tomography
for their determination and for which of them such a complete
knowledge is redundant. In this paper we analyze the resources
needed for a comparison of quantum states. Suppose a given pair of
quantum systems in unknown states. The question is what
experiments (if any) are capable either of revealing with
certainty the difference between the states, or confirming their
sameness as long as the the probability distribution of
measurement outcomes is identified.

By the very nature of quantum theory, the events we observe in
quantum experiments are random. That is, both quantum predictions
and quantum conclusions are naturally formulated in terms of
probabilities and uncertainty. Therefore, it is surprising that
there are (very specific) situations (including special instances
of the comparison problem) in which individual clicks enable us to
make a nontrivial unambiguous prediction, or conclusion. For
example, if we are given a promise that the states are pure, then
(with a nonzero probability) the difference of states can be
confirmed unambiguously from a single experimental
click~\cite{Barnett03,Andersson06}. This result can be also
generalized to the comparison of many pure
states~\cite{jex-jmo-2004,chefles-jpa-2004,Kleinmann-2005}, the
comparison of ensembles of pure states~\cite{sedlak-pra-2008}, and
the comparison of some pure continuous-variable
states~\cite{sedlak-pra-2007,Olivares-pra-2011} (see also the
review~\cite{Sedlak09_aps}). Unfortunately, such single-shot
(non-statistical) comparison strategy fails for general mixed
states~\cite{Kleinmann-2005,Pang11}. The reason is simple. The
probability of any outcome is strictly nonvanishing provided that
a bipartite system is in the completely mixed state, for which the
subsystems are in the same state. That is, for any outcome there
is a situation in which the systems are the same, hence the
difference cannot be concluded unambiguously. In such a case any
error-free conclusions need to be based on the observed
probabilities of outcomes. Probability-based strategies were not
considered in previous studies of quantum state comparison. Our
aim in this paper is to introduce this concept and provide basic
results in this area.

Trivially, if the experimentally measured probabilities provide
complete information on quantum states of both systems
individually, then they also contain all the information needed
for the comparison. The question of our interest is whether the
complete tomography is necessary.  Our main goal is to design a
comparison experiment providing as little redundant information as
possible.

In Sec.~\ref{section:problem}, we introduce the necessary
mathematical notation and formulate the problem. In
Sec.~\ref{section:universal}, we address the existence of a
universal comparison measurement. Sec.~\ref{section:two-valued}
investigates the comparison performance of two-outcome
measurements. Almost universal two-valued and many-valued
comparison measurements are presented in
Sec.~\ref{section:AUC-measurement} and conclusions are the content
of Sec.~\ref{section:summary}.

%-------------------------------------------------------------------------------------------

\section{\label{section:problem} Problem formulation}
Any quantum state is associated with the density operator
$\varrho\in\cS(\cH)$ such that $\varrho\geq O$ and
$\tr{\varrho}=1$. Hereafter, $\cS(\cH)$ stands for the set of all
states of a system associated with the Hilbert space $\cH$. The
statistical features of quantum measurements are fully captured by
means of a positive operator-valued measure (POVM) that is a
collection $\E$ of positive operators (acting on $\cH$ and called
effects) $E_1,\dots,E_n$ summing up to the identity, i.e.
$\sum_{j=1}^n E_j=I$. For each state $\varrho\in\cS(\cH)$ the
measurement $\E$ assigns a probability distribution
$\{p_j\}_{j=1}^{n} \equiv \vec{p}_{\E}$, where $p_j=\tr{E_j
\varrho} \ge 0$ and $\sum_{j=1}^n p_j =1$.

Let us now move on to the set of bipartite
factorized states $\cS_{\rm fac} = \{\varrho\otimes\xi :
\varrho,\xi\in\cS(\cH)\} \subset \cS(\cH\otimes\cH)$, where the
parties $\varrho$ and $\xi$ are the states to be compared. For a
fixed measurement $\E$ we can ask how much information it reveals
concerning the comparison of the subsystems.

\begin{figure}
\includegraphics[width=8cm]{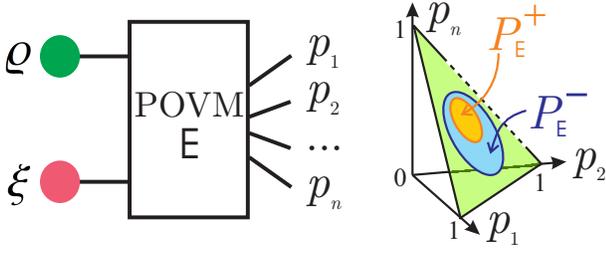}
\caption{\label{fig:prob-comp} (Color online) Illustration of
probability-based comparison. If the observed probability
distribution belongs to $P_{\E}^- \setminus P_{\E}^+$, then states
$\varrho$ and $\xi$ are for sure different.}
\end{figure}

Denote by $\cS^+$ the subset of \emph{twin-identical states}, i.e.
$\cS^+=\{\eta\otimes\eta: \eta\in\cS(\cH)\}\subset
\cS(\cH\otimes\cH)$. Similarly, let us denote by $\cS^-$ the
subset of \emph{non-identical states}, i.e.
$\cS^-=\{\varrho\otimes\xi:\varrho,\xi(\neq\varrho)\in\cS(\cH)\}\subset
\cS(\cH\otimes\cH)$. Obviously, $\cS_{\rm fac}=\cS^+\cup\cS^-$.
The goal of comparison is then to distinguish between sets of
states $\cS^+$ and $\cS^-$. This goal can be achieved in our
approach by considering two sets of probability distributions
$P_{\E}^{\pm} = \{ \vec{p}: p_j = \tr{E_j \omega} , \omega \in
\cS^{\pm} \}$. In other words, since the measurement $\E$ performs
the mapping $\cS^{\pm} \mapsto P_{\E}^{\pm}$, one can
unambiguously conclude that a bipartite state $\omega$ belongs to
the set $\cS^{\pm}$ if the observed probability distribution
$\vec{p}_{\E} \in P_{\E}^{\pm} \setminus P_{\E}^{\mp}$ (see
Fig.~\ref{fig:prob-comp}).

%%%%%%%%%%%%%%%%%%%%%%%%%%%%%%%%%%%%%%%%%%%%%%%%%%%%%%%%%%%%%%%%%%%
%%%%%%%%%%%%%%%%%%%%%%%%%%%%%%%%%%%%%%%%%%%%%%%%%%%%%%%%%%%%%%%%%%%
%%%%%%%%%%%%%%%%%%%%%%%%%%%%%%%%%%%%%%%%%%%%%%%%%%%%%%%%%%%%%%%%%%%
%%%%%%%%%%%%%%%%%%%%%%%%%%%%%%%%%%%%%%%%%%%%%%%%%%%%%%%%%%%%%%%%%%%

For a fixed POVM $\E$ on $\cH\otimes\cH$ we may introduce the
following quantities:
\begin{eqnarray}
\label{eq:dist_1} \!\!\!\!\! D_\E(\varrho\otimes\xi,\cS^+) &=&
\inf\limits_{\eta\otimes\eta \in \cS^+}
\sum_{j=1}^n \left| \tr{E_j(\varrho\otimes\xi - \eta\otimes\eta)} \right|, \\
\label{eq:dist_2} \!\!\!\!\! D_\E(\eta\otimes\eta,\cS^-) &=&
\inf\limits_{\varrho\otimes\xi \in \cS^-} \sum_{j=1}^n \left|
\tr{E_j(\varrho\otimes\xi - \eta\otimes\eta)} \right|.
\end{eqnarray}

While $D_\E(\varrho\otimes\xi,\cS^+)$ quantifies how different the
states $\varrho$ and $\xi$ are (with respect to measurement $\E$),
the value of $D_\E(\eta\otimes\eta,\cS^-)$ tells us to which
extent the equivalence of twin-identical states can be confirmed.

Before we proceed further let us make one important observation:
for all $\epsilon>0$ and any state $\eta\otimes\eta$ there exists
a state $\varrho\otimes\xi$ such that
$|\tr{E(\eta\otimes\eta-\varrho\otimes\xi)}| \le \epsilon$ for any
POVM effect $E$. In other words, in order to conclude that the
states are the same no uncertainty in the specification of the
probabilities $p_E(\omega)=\tr{E \omega}$ is allowed. Such an
infinite precision is practically not achievable, however, for our
purposes we will assume the probabilities are specified exactly.
The proof of the statement above is relatively straightforward.
Let us set $\varrho=\eta$ and
$\xi=(1-\frac{\epsilon}{2})\eta+\frac{\epsilon}{2d}I$, i.e.
$\eta\otimes\eta-\varrho\otimes\xi=\frac{\epsilon}{2}
\eta\otimes(\eta-\frac{1}{d}I)$. Since $|\tr{EX}| \le
\max_{E\in\E} \tr{|EX|} \le \tr{|X|}$, it follows that
\begin{equation}
\label{inequality} \left|
\tr{E(\eta\otimes\eta-\varrho\otimes\xi)} \right| \le \epsilon \,
\tfrac{1}{2} \tr{|\eta\otimes(\eta-\tfrac{1}{d}I)|} \le
\epsilon\,.
\end{equation}

\noindent In the last inequality we used the fact that the trace
distance of states is bounded from above by one. Formula
(\ref{inequality}) is valid for any POVM-effect $E_j$, therefore
$\sum_{j=1}^n \left| \tr{E_j(\varrho\otimes\xi - \eta\otimes\eta)}
\right| \le n \epsilon \rightarrow 0$ when $\epsilon \rightarrow
0$. By definition of the greatest lower bound (infimum), the
distance (\ref{eq:dist_2}) vanishes for an arbitrary
$\eta\in\cS(\cH)$.

Our observation implies that for any measurement $\E$ we have
$D_\E(\eta\otimes\eta,\cS^-)=0$. This seems to be in contradiction
with measurements which provide us with complete information on
the states of individual systems. We will refer to such
measurements as \emph{locally informationally complete} (LIC)
measurements. Clearly, in case of an ideal LIC measurement the
sameness can be verified. Where is the problem? Topologically, in
the set of factorized states $\cS_{\rm fac}$ with the
trace-distance metrics, the subset $\cS^+$ is closed and does not
contain any interior point (therefore the distance
(\ref{eq:dist_2}) vanishes), however, it does not mean that the
subset $\cS^+$ is empty.

\textit{Remark}. The subset $\cS^+$ is closed because the set
$\cS(\cH)$ is closed. To prove that $\cS^+$ does not contain any
interior point, assume the converse. Let $\eta_0\otimes\eta_0$ be
an interior point of $\cS^{+}$, then there exists a neighborhood
$\cO_{\varepsilon}(\eta_0\otimes\eta_0)$ such that
$\cO_{\varepsilon}(\eta_0\otimes\eta_0) \subset \cS^+$. Choose an
arbitrary point
$\eta\otimes\eta\in\cO_{\varepsilon}(\eta_0\otimes\eta_0)$ with
$\eta\ne\eta_0$, then a nontrivial convex combination $[\lambda
\eta_0\otimes\eta_0 + (1-\lambda) \eta\otimes\eta] \in
\cO_{\varepsilon}(\eta_0\otimes\eta_0) \subset \cS^+$, i.e.
$\lambda \eta_0\otimes\eta_0 + (1-\lambda) \eta\otimes\eta =
\zeta\otimes\zeta$ for some $\zeta\in\cS(\cH)$. Taking partial
trace over the first subsystem, we obtain
$\lambda\eta_0+(1-\lambda)\eta = \zeta$. In view of this,
$\zeta\otimes\zeta = \lambda^2 \eta_0\otimes\eta_0 +
\lambda(1-\lambda)(\eta_0\otimes\eta+\eta\otimes\eta_0)+(1-\lambda)^2
\eta\otimes\eta$. Subtracting the two expressions obtained for
$\zeta\otimes\zeta$ yields $(\eta-\eta_0)\otimes(\eta-\eta_0)=0$,
i.e. $\eta=\eta_0$, which contradicts the choice $\eta \ne
\eta_0$. Thus, $\cS^+$ does not contain any interior point.

The vanishing value of the distance considered (\ref{eq:dist_2})
is not completely relevant if one thinks about the ideal
error-free experiments. In practice, experimental noise is
unavoidable; hence, from the practical point of view a conclusion
on the sameness of states can never be error free.

%-------------------------------------------------------------------------------------------

\section{\label{section:universal} Universal comparison measurement}

We say the measurement $\E$ implements the comparison whenever
$D_\E(\varrho\otimes\xi,S^+)>0$ for some pairs $\varrho,\xi$. The
state comparison measurement $\E$ is \emph{universal} if
$D_\E(\varrho\otimes\xi,S^+)>0$ for all
$\varrho,\xi(\neq\varrho)$. This is, for instance, achieved in
case of the ideal LIC measurements: even though the value of
$D_\E(\varrho\otimes\xi,S^+)$ can be arbitrarily small, it always
remains strictly positive. As before, this situation is not very
realistic in practice, because any error in the identification of
outcome probabilities makes the conclusions (in some cases of
$\varrho$ and $\xi$) ambiguous. However, assuming the infinite
precision in the specification of probabilities, the universality
can be achieved and in what follows we will assume that
probabilities are identified perfectly. The potential errors can
be viewed as modifications of the sets $\cS^+$ and $\cS^-$ we are
aiming to distinguish. Nevertheless, our goal is to analyze the
ideal case.

Let us now demonstrate that a universal comparison can be
implemented if and only if the measurement is LIC.

To start with, we are reminded that any POVM $\E$ with effects
$E_j$ linearly maps a state $\omega \in \cS(\cH\otimes\cH)$ into
the probability vector $\vec{p} = (p_1,p_2,\ldots)$, where $p_j =
\tr{E_j \omega}$. For LIC measurements the induced mapping
$\varrho\otimes\xi\mapsto\vec{\pi}$ is bijective. That proves the
sufficiency. To prove the necessity let us assume the converse,
i.e. suppose the measurement $\E$ is not an LIC measurement but
implements a universal comparison. Since $\E$ is not LIC, the
probability assignment $\varrho\otimes\xi\mapsto\vec{p}_{\E}$ is
injective. Let us denote by $\Pi^{\pm}$ and $\Pi$ the images of
$\cS^{\pm}$ and $\cS_{\rm fac}$ under some LIC measurement,
respectively, and by $P_{\E}$ denote the image of $\cS_{\rm fac}$
under the measurement $\E$. Clearly, the relation between
$\vec{\pi}(\varrho\otimes\xi)$ and
$\vec{p}_{\E}(\varrho\otimes\xi)$ is linear and injective, i.e.
there exist probability vectors $\vec{\pi}_1\in\Pi$ and
$\vec{\pi}_2\in\Pi$ transformed into the same probability
distribution $\vec{p}_{\E}(\varrho\otimes\xi)\in P_{\E}$. The
distributions $\vec{\pi}_j$ transformed into the same probability
vector $\vec{p}_{\E}(\varrho\otimes\xi)$ span a linear subspace
(hyperplane) $H_{\varrho\otimes\xi}$ in the linear span of $\Pi$.

%%%%%%%%%%%%%%%%%%%%%%%%%%%%%%%%%%%%%%%%%%%%%%%%%%%%%%%%%%%%%%%%%%%
%%%%%%%%%%%%%%%%%%%%%%%%%%%%%%%%%%%%%%%%%%%%%%%%%%%%%%%%%%%%%%%%%%%
\begin{figure*}
\includegraphics[width=18cm]{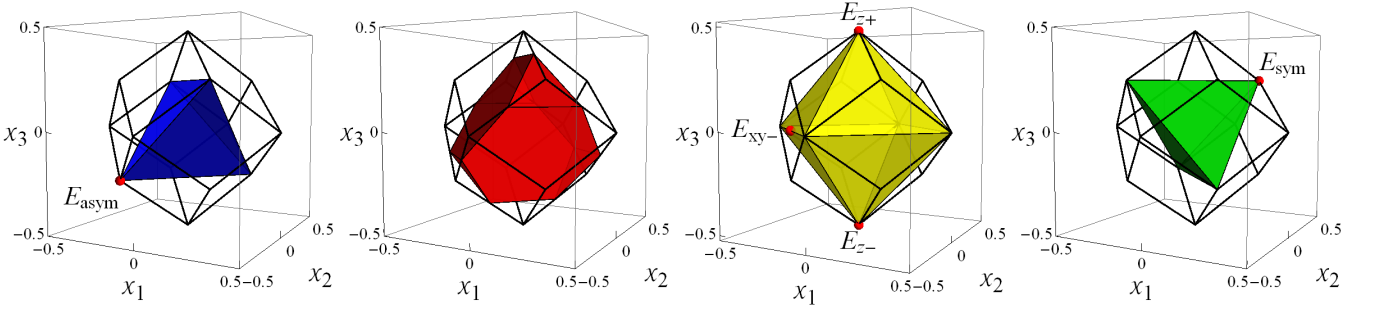}
\caption{\label{figure:Rhombo} (Color online) Body
$B(\varkappa_0)$, i.e. the region of parameters
$(\varkappa_1,\varkappa_2,\varkappa_3)$ when (\ref{E-diag}) is a
true POVM effect. Parameter $\varkappa_0$ takes values
$\frac{1}{4}$, $\frac{3}{8}$, $\frac{1}{2}$, and $\frac{3}{4}$ for
figures from left to right. The union $\cup_{\varkappa_0 \in
[0,1]} B(\varkappa_0)$ is a rhombododecahedron and is depicted by
solid lines. POVM effects $E_{\rm asym}$, $E_{\rm sym}$, $E_{z
\pm}$ are vertices and POVM effects $E_{xy \pm}$ are face centers
of this convex polytope.}
\end{figure*}
%%%%%%%%%%%%%%%%%%%%%%%%%%%%%%%%%%%%%%%%%%%%%%%%%%%%%%%%%%%%%%%%%%%
%%%%%%%%%%%%%%%%%%%%%%%%%%%%%%%%%%%%%%%%%%%%%%%%%%%%%%%%%%%%%%%%%%%

Consider an internal point $\eta\in\cS(\cH)$, then the image
$\vec{\pi}(\eta\otimes\eta)$ is an interior point of $\Pi$ on the
probability simplex. There exists $\epsilon_0>0$ such that for all
$0<\epsilon\leq \epsilon_0$ the neighborhood
$O_\epsilon(\vec{\pi}(\eta\otimes\eta))$ belongs to $\Pi$ (on the
simplex). Moreover, the intersection $O_\epsilon
(\vec{\pi}(\eta\otimes\eta))\cap H_{\eta\otimes\eta}$ cannot be a
subset of $\Pi^+$ only, because $\Pi^+$ does not contain any
interior point on the simplex (if it did, the distance
$D_{\E}(\eta\otimes\eta,\cS^-)$ would not vanish for all states
$\eta$). Thus, $O_\epsilon (\vec{\pi}(\eta\otimes\eta)) \cap
H_{\eta\otimes\eta} \cap \Pi^-$ is not empty and contains points
of the form $\vec{\pi}(\tilde{\varrho}\otimes\tilde{\xi})$ such
that $\tilde{\varrho} \neq \tilde{\xi}$. As both
$\vec{\pi}(\eta\otimes\eta)$ and
$\vec{\pi}(\tilde{\varrho}\otimes\tilde{\xi})$ belong to
$H_{\eta\otimes\eta}$, we have
$\vec{p}_{\E}(\tilde{\varrho}\otimes\tilde{\xi}) =
\vec{p}_{\E}(\eta\otimes\eta)$ and formula (\ref{eq:dist_1})
yields $D(\tilde{\varrho}\otimes\tilde{\xi},\cS^+)=0$, i.e. $\E$
is not a universal comparison measurement (by definition). This
contradiction concludes the proof of the necessity.

Let us summarize two main conclusions:

(i) In any locally informationally incomplete measurement the
sameness of states cannot be confirmed.

(ii) Universal comparison (concluding universally and
unambiguously the difference of states) requires a locally
informationally complete measurement.

A question that remains open is how to evaluate the overall
performance of (universal or non-universal) comparison
experiments. There are several options. We can use the volume of
the subset $\cS_{\rm comp}^-$ of states in $\cS^-$ that can be
successfully compared, or the average value of
$D_\E(\varrho\otimes\xi,\cS^+)$ with respect to some measure on
the state space. In particular, these quantities read
\begin{eqnarray}
\label{S-comp} |\cS^-_{\rm comp}|_\E &=& \iint_{\cS^-}
\mu(d\varrho)\mu(d\xi) h(D_\E(\varrho\otimes\xi,\cS^+))\,,\\
\label{D-E-ave} \ave{D_\E} &=& \iint_{\cS^-}
\mu(d\varrho)\mu(d\xi) D_\E(\varrho\otimes\xi,\cS^+)\,,
\end{eqnarray}

\noindent where $h(x)$ is the Heaviside function and
$\mu(d\varrho)=\mu(d\xi)$ is a measure on the state space of
individual subsystems. Quite common choices for the measure $\mu$
on density operators are the ones induced by metrics, namely, by
Bures distance and Hilbert--Schmidt distance (see,
e.g.,~\cite{Hall,Bengtsson} and references therein). Let us stress
that $|\cS^-_{\rm comp}|_\E=1$ does not imply the comparison is
universal, because there can be a set of measure zero for which
$D_\E(\varrho\otimes\xi,S^+)=0$. In such case we say that the
comparison measurement is \emph{almost universal}. It is of great
interest to investigate whether there exist some almost universal
comparison experiments and, in particular, how many outcomes such
measurements require.

%-------------------------------------------------------------------------------------------

\section{\label{section:two-valued} Two-valued comparison experiments}
Let us start our investigation with the simplest case of
two-valued POVMs described by the effects $E$ and $I-E$.  In such
a case,
\begin{equation}
\label{distance-two-valued}
D_\E(\varrho\otimes\xi,\cS^+)=2D_E(\varrho\otimes\xi,\cS^+)\,,
\end{equation}

\noindent where
$D_E(\varrho\otimes\xi,\cS^+)=\min_{\eta\otimes\eta\in\cS^+}|\tr{E(\varrho\otimes\xi-
\eta\otimes\eta)}|$. Two-valued measurements cannot be LIC,
because they provide the only informative real number (the
probability $p_E$, $p_{I-E} = 1-p_E$) whereas the state
$\varrho\otimes\xi$ is defined by $2(d^2-1)\ge 6$ real numbers.
Thus, two-valued measurements are necessarily non-universal
comparators. Nevertheless, it is of practical interest to
understand how good their comparison performance is.

Let us consider a geometry of comparable states. Suppose
$\varrho=\frac{1}{d}(I+{\bf r}\cdot\boldsymbol{\Lambda})$ and
$\xi=\frac{1}{d}(I+{\bf k}\cdot\boldsymbol{\Lambda})$, where
$\boldsymbol{\Lambda}=(\Lambda_1,\dots,\Lambda_{d^2-1})$ is a
vector formed of traceless Hermitian operators $\Lambda_j$ such
that $\tr{\Lambda_j\Lambda_k}=d\delta_{jk}$, and ${\bf r},{\bf k}
\in \mathbb{R}^{d^2-1}$ are Bloch-like vectors which necessarily
satisfy $|{\bf r}|,|{\bf k}|\le \sqrt{d-1}$ (see,
e.g.,~\cite{heinosaari-ziman}). Using this notation, let us find
such vectors ${\bf r}$ that the states $\varrho$ are comparable
with a fixed state $\xi$ (the POVM-effect $E$ is fixed as well).
The trace $\tr{E\varrho\otimes\xi} = \frac{1}{d} (\tr{E
I\otimes\xi} + {\bf r}\cdot {\bf K})$, where ${\bf K} =
\tr{E\boldsymbol\Lambda\otimes\xi}$. Therefore the inequality
$D_E(\varrho\otimes\xi,\cS^+)>0$ boils down to either
\begin{equation}
\label{rK-greater} \tfrac{1}{d} \, {\bf r}\cdot {\bf K} >
\max_{\eta\otimes\eta\in\cS^{+}} \tr{E\eta\otimes\eta} -
\tfrac{1}{d} \, \tr{E I\otimes\xi},
\end{equation}

\noindent or
\begin{equation}
\label{rK-smaller} \tfrac{1}{d} \, {\bf r}\cdot {\bf K} <
\min_{\eta\otimes\eta\in\cS^{+}} \tr{E\eta\otimes\eta} -
\tfrac{1}{d} \, \tr{E I\otimes\xi}.
\end{equation}

\noindent These inequalities define two nonintersecting
half-spaces in $\mathbb{R}^{d^2-1}$ separated by the distance
\begin{equation}
\label{L-distance} L=\frac{d}{|{\bf K}|} \left(
\max_{\eta\otimes\eta\in\cS^{+}} \tr{E\eta\otimes\eta} -
\min_{\eta\otimes\eta\in\cS^{+}} \tr{E\eta\otimes\eta} \right).
\end{equation}

\noindent Thus, for any fixed $\xi$ the set of successfully
comparable Bloch-like vectors ${\bf r}$ is given by an
intersection of two half-spaces with the state space.

\subsection{SWAP-based comparison}
As we have already mentioned in Sec.~\ref{section:introduction},
if we restrict ourselves only to pure states, then there exists a
strategy to perform an unambiguous comparison (via the SWAP
measurement). In such an approach, the sameness of the states
cannot be concluded and this is related to the absence of the
universal NOT-operation~\cite{barnett-jmo-2010}. However, the
strategy (if successful) can reveal the difference between the
states in a single shot, hence, no collection of statistics is
needed.

The key observation for such a conventional strategy is that the
support of twin-identical pure states spans only the symmetric
subspace of $\cH\otimes\cH$. Suppose projections $E_{\rm
sym},E_{\rm asym}$ onto symmetric and antisymmetric subspaces of
$\cH\otimes\cH$. Since $E_{\rm sym}+E_{\rm asym}=I$ they form a
two-valued POVM $\E_{\rm SWAP}$. Let us note that $E_{\rm
sym}=\frac{1}{2}(I+S)$, $E_{\rm asym}=\frac{1}{2}(I-S)$, where $S$
is the SWAP operator acting as
$S(\ket{\psi\otimes\varphi})=\ket{\varphi\otimes\psi}$ for all
$\ket{\psi},\ket{\varphi}\in\cH$. It is straightforward to see
that for any twin-identical pure state
$\ket{\varphi\otimes\varphi}$ one has $\tr{E_{\rm asym}
\ket{\varphi\otimes\varphi}\bra{\varphi\otimes\varphi} }=0$,
however $\tr{E_{\rm asym}
\ket{\varphi\otimes\psi}\bra{\varphi\otimes\psi} } =
\frac{1}{2}(1-|\ip{\varphi}{\psi}|^2) > 0$ if $\ket{\psi} \neq
\ket{\varphi}$. Therefore, recording an outcome $E_{\rm asym}$
allows us to unambiguously conclude that the states are different.
No statistics is needed.

Let us see how this  strategy works in the case of general mixed
states. A direct calculation yields
\begin{eqnarray}
p_{\rm sym}&=&\tr{E_{\rm sym}\varrho\otimes\xi}=\tfrac{1}{2}(1+\tr{\varrho\xi}) \, ,\\
p_{\rm asym}&=&\tr{E_{\rm
asym}\varrho\otimes\xi}=\tfrac{1}{2}(1-\tr{\varrho\xi}) \, ,
\end{eqnarray}

\noindent where we used the identity
$\tr{S\varrho\otimes\xi}=\tr{\varrho\xi}$. The purity of a state,
$\tr{\eta^2}$, is bounded from below by $1/d$, where $d = \dim
\cH$. Thus,
\begin{eqnarray}
&& p_{\rm sym}(\varrho\otimes\xi)\in [\tfrac{1}{2},1) \equiv P_{\rm sym}^-\,,\\
&& p_{\rm sym}(\eta\otimes\eta)\in[\tfrac{d+1}{2d},1] \equiv
P_{\rm sym}^+\,,
\end{eqnarray}

\noindent where $P_{\rm sym}^{\pm}$ is the image of $\cS^{\pm}$
under the POVM effect $E_{\rm sym}$. It follows that by measuring
the probability $p_{\rm sym}<(d+1)/2d$ we can with certainty
conclude that the states are different. In particular, $D_{\rm
SWAP}(\varrho\otimes\xi,\cS^+)=\max\{0, \frac{1}{d}
-\tr{\varrho\xi}\}$.

Associating $\varrho$ and $\xi$ with the Bloch-like vectors ${\bf
r},{\bf k} \in \mathbb{R}^{d^2-1}$ as above ($|{\bf r}|,|{\bf
k}|\le \sqrt{d-1}$), for the SWAP-based measurement we obtain
${\bf K}_{\pm} = \tr{(I \pm S)\boldsymbol\Lambda\otimes\xi} = \pm
{\bf k}$. Also, we find explicitly $\tr{(I\pm S) I\otimes\xi}=d
\pm 1$, $\max_{\eta\otimes\eta\in\cS^{+}} \tr{(I+S)
\eta\otimes\eta} = 2$, $\max_{\eta\otimes\eta\in\cS^{+}} \tr{(I-S)
\eta\otimes\eta} = 1-\tfrac{1}{d}$,
$\min_{\eta\otimes\eta\in\cS^{+}} \tr{(I+S) \eta\otimes\eta} =
1+\tfrac{1}{d}$, and $\min_{\eta\otimes\eta\in\cS^{+}} \tr{(I-S)
\eta\otimes\eta} = 0$. Then it is straightforward to see that, for
the SWAP-based measurement, one of inequalities (\ref{rK-greater})
and (\ref{rK-smaller}) is never fulfilled and the other one
reduces to ${\bf r}\cdot{\bf k}<0$. That is, for each fixed ${\bf
k}(\neq\boldsymbol{0})$ the set of successfully comparable
Bloch-like vectors ${\bf r}$ is given by an intersection of a
single half-space with the state space. Let us stress that for
qubits ($d=2$) the state space is exactly the Bloch ball $|{\bf
r}|\le 1$, so the set of comparable vectors ${\bf r}$ is the
hemisphere. Thus, for qubits $|\cS^-_{\rm comp}|_{\rm
SWAP}=\frac{1}{2}$ (see the next paragraph). In other words, the
difference of states from the same hemisphere is not detected in
the SWAP measurement. This implies that the approximate
universality is lost.

Due to unitary invariance of the measure $\mu$ we can always treat
one of the states in $D_E(\varrho\otimes\xi,\cS^+)$ as diagonal,
say $\varrho$. Then $\tr{\varrho\xi}=\sum_{j=1}^d
\varrho_{jj}\xi_{jj}$ and the integration area of $|\cS^-_{\rm
comp}|_{\rm SWAP}$ is split into $d!$ subsets labeled by the
permutation of the labels $j_1,\dots,j_d$ identifying the ordering
$\varrho_{j_1 j_1}\geq\dots\geq\varrho_{j_d j_d}$ of eigenvalues
of $\varrho$. The (normalized) volume of each of these subsets is
$\frac{1}{d!}$. If $\varrho$ and $\xi$ are from mutually opposite
subsets (labeled as $j_1,\dots,j_d$ and $j_d,\dots, j_1$,
respectively), then $\tr{\varrho\xi}=\sum_{j=1}^d
\varrho_{jj}\xi_{jj} \leq \frac{1}{d}$ (see the remark below)
meaning that such pairs $\varrho$ and $\xi$ can be successfully
compared. Therefore, we find the lower bound $|\cS^-_{\rm
comp}|_{\rm SWAP}\geq \frac{1}{d!}$. If $\varrho,\xi$ are from the
same subset, then $\tr{\varrho\xi}=\sum_{j=1}^d
\varrho_{jj}\xi_{jj}\geq \frac{1}{d}$ (see the remark below),
hence the contribution to $|\cS^-_{\rm comp}|_{\rm SWAP}$ is
vanishing and we can bound the fraction of comparable states also
from above, $|\cS^-_{\rm comp}|_{\rm SWAP} \leq 1-\frac{1}{d!}$.
For qubits both bounds coincide and give the value $|\cS^-_{\rm
comp}|_{\rm SWAP}=\frac{1}{2}$.

\textit{Remark}. The $d$-dimensional probability vectors
$\boldsymbol\varrho = (\varrho_{j_1 j_1},\ldots,\varrho_{j_d
j_d})$ obeying the ordering $\varrho_{j_1
j_1}\geq\dots\geq\varrho_{j_d j_d}$ form a convex subset on
($d-1$)-simplex, with the extremal points being
$(1,0,0,\ldots,0)$, $(\frac{1}{2},\frac{1}{2},0,\ldots,0)$,
\ldots,
$(\frac{1}{d},\frac{1}{d},\frac{1}{d},\ldots,\frac{1}{d})$.
Analogously, the $d$-dimensional probability vectors
$\boldsymbol\xi =(\xi_{j_1 j_1},\ldots,\xi_{j_d j_d})$ obeying the
ordering $\xi_{j_1 j_1}\leq\dots\leq\xi_{j_d j_d}$ form a convex
subset on ($d-1$)-simplex, with the extremal points being
$(0,\ldots,0,0,1)$, $(0,\ldots,0,\frac{1}{2},\frac{1}{2})$,
\ldots,
$(\frac{1}{d},\ldots,\frac{1}{d},\frac{1}{d},\frac{1}{d})$. Since
the function $f(\boldsymbol\varrho,\boldsymbol\xi) =
\boldsymbol\varrho \cdot \boldsymbol\xi =\sum_{j=1}^d
\varrho_{jj}\xi_{jj} = \tr{\varrho\xi}$ is linear with respect to
$\varrho_{jj}$ and $\xi_{jj}$, then its extremal (maximum) value
is achieved for some pair $(\boldsymbol\varrho,\boldsymbol\xi)$ of
extremal probability vectors $\boldsymbol\varrho$ and
$\boldsymbol\xi$. It can be easily checked that for all pairs of
extremal probability vectors $\boldsymbol\varrho \cdot
\boldsymbol\xi \le \frac{1}{d}$, i.e.
$\tr{\varrho\xi}\le\frac{1}{d}$. Similarly, for the ordering
$\xi_{j_1 j_1}\geq\dots\geq\xi_{j_d j_d}$ one has
$\tr{\varrho\xi}\ge\frac{1}{d}$.

%-------------------------------------------------------------------------------------------

\subsection{Qubit ``diagonal" comparison experiments}
In the previous subsection we have shown that the SWAP-based
comparison enables us (in the case of qubits) to successfully
detect the difference for half (up to a set of zero measure) of
the pairs $\varrho\otimes\xi$. Can one do better with some other
two-valued measurement?

Arguing as in the beginning of Sec.~\ref{section:two-valued}, we
can conclude that, for a given effect $E$, the states $\varrho$
comparable with a fixed state $\xi$ form a cut of the Bloch ball
or two cuts of this ball by parallel planes. Note that for the
SWAP-based comparison the complete mixture cannot be conclusively
compared with any other state. So we will be interested in finding
such a two-valued measurement, for which the complete mixture can
be unambiguously distinguished from some other states.

A general two-qubit effect takes the form
\begin{equation}
E=\textstyle\sum_{l,m=0}^{3}\varepsilon_{lm}\sigma_l \otimes
\sigma_m,
\end{equation}

\noindent where we use the notation $\sigma_0=I$ and
$\sigma_1,\sigma_2,\sigma_3$ denote the Pauli operators
$\sigma_x,\sigma_y,\sigma_z$, respectively. Real coefficients
$\varepsilon_{lm}$ read $\varepsilon_{lm} = \frac{1}{4}{\rm tr}[E
\sigma_l \otimes \sigma_m]$ and satisfy the the constraints $O\leq
E\leq I$.

For the sake of simplicity let us consider only the diagonal case,
i.e. we will assume $\varepsilon_{lm}=\delta_{lm}\varkappa_m$,
with $\varkappa_m \in \mathbb{R}$ for all $m=0,\dots,3$. Then
\begin{equation}
\label{E-diag}
E_{\rm diag} = \left(%
\begin{array}{cccc}
  \varkappa_0+\varkappa_3 & 0 & 0 & \varkappa_1-\varkappa_2 \\
  0 & \varkappa_0-\varkappa_3 & \varkappa_1+\varkappa_2 & 0 \\
  0 & \varkappa_1+\varkappa_2 & \varkappa_0-\varkappa_3 & 0 \\
  \varkappa_1-\varkappa_2 & 0 & 0 & \varkappa_0+\varkappa_3 \\
\end{array}%
\right).
\end{equation}

\noindent Applying the positivity constraints on $E$ of this form
we obtain the following conditions:
\begin{eqnarray}
\label{inequal-lambda1} 0 \le \varkappa_0 - \varkappa_1 -
\varkappa_2 -
\varkappa_3 \le 1, \\
0 \le \varkappa_0 - \varkappa_1 + \varkappa_2 +
\varkappa_3 \le 1, \\
0 \le \varkappa_0 + \varkappa_1 - \varkappa_2 +
\varkappa_3 \le 1, \\
\label{inequal-lambda4} 0 \le \varkappa_0 + \varkappa_1 +
\varkappa_2 - \varkappa_3 \le 1.
\end{eqnarray}

\noindent The system of inequalities
(\ref{inequal-lambda1})--(\ref{inequal-lambda4}) has a nontrivial
solution whenever $0<\varkappa_0<1$. Indeed, if $\varkappa_0$ is
fixed, then each two-sided inequality of the system determines a
geometric fiber between two planes in the reference frame
$(\varkappa_1,\varkappa_2,\varkappa_3)$, with the distance between
the planes being equal to $\frac{1}{\sqrt{3}}$. If
$0<\varkappa_0\le \frac{1}{4}$, then four fibers intersect to
yield a tetrahedron. The intersection becomes a truncated
tetrahedron if $\frac{1}{4} < \varkappa_0 < \frac{1}{2}$ and
finally transforms into an octahedron for the case
$\varkappa_0=\frac{1}{2}$. If $\varkappa_0 \ge \frac{1}{2}$ then
the solution is a body obtained by the inversion
$(\varkappa_1,\varkappa_2,\varkappa_3) \rightarrow
(-\varkappa_1,-\varkappa_2,-\varkappa_3)$ of the body labeled by
parameter $(1-\varkappa_0)$. For instance, if $\frac{3}{4} \le
\varkappa_0 < 1$, then the intersection is a tetrahedron inverted
with respect to that in the case $0<\varkappa_0\le \frac{1}{4}$.
Given $\varkappa_0$ we will refer to the intersection as body
$B(\varkappa_0)$ (see Fig.~\ref{figure:Rhombo}).

The probabilities of the measurement outcome, corresponding to the
POVM effect $E_{\rm diag}$, can be readily calculated for the
non-identical (different) and twin-identical (same) states and
read
\begin{eqnarray} \label{p-diff} p_{\rm diff} &=&
p_{\rm diag}(\varrho\otimes\xi)= \varkappa_0 +
\textstyle\sum_{m=1}^3\varkappa_m r_m
k_m \,,\\
\label{p-same} p_{\rm same} &=& p_{\rm diag}(\eta\otimes\eta)=
\varkappa_0 + \textstyle\sum_{m=1}^3\varkappa_m h_m^2 \,,
\end{eqnarray}

\noindent where we used $\varrho=\frac{1}{2}(I+{\bf
r}\cdot\boldsymbol{\sigma})$, $\xi=\frac{1}{2}(I+{\bf
k}\cdot\boldsymbol{\sigma})$ and $\eta=\frac{1}{2}(I+{\bf
h}\cdot\boldsymbol{\sigma})$. Using the normalization constraints
for Bloch vectors ${\bf r}$, ${\bf k}$, and ${\bf h}$, we obtain
from Eqs. (\ref{p-diff}) and (\ref{p-same}) that probabilities
$p_{\rm diff}$ and $p_{\rm same}$ satisfy the relations
\begin{eqnarray}
&& p_{\rm diff} \in [\varkappa_0-\varkappa_{\max}, \varkappa_0+\varkappa_{\max}] \equiv {\rm cl}(P_{\rm diag}^-) \,,\\
&& p_{\rm same} \in [\varkappa_0-|\varkappa_-|,
\varkappa_0+\varkappa_+] \equiv P_{\rm diag}^+\,,
\end{eqnarray}

\noindent respectively, where we introduced the notations
$\varkappa_{\max}=\max\{|\varkappa_1|,|\varkappa_2|,|\varkappa_3|\}$,
$\varkappa_-=\min\{0,\varkappa_1,\varkappa_2,\varkappa_3\}$, and
$\varkappa_+=\max\{0,\varkappa_1,\varkappa_2,\varkappa_3\}$.
Clearly, $P_{\rm diag}^+ \subset {\rm cl}(P_{\rm diag}^-)$, hence,
there is a two-valued POVM that allows us to make a nontrivial
conclusion about the difference of some states. Notice that the
case $\varkappa_0=\frac{3}{4}$,
$\varkappa_1=\varkappa_2=\varkappa_3=\frac{1}{4}$ gives the
SWAP-based comparison measurement for which, whenever the measured
probability $p_{\rm diag}$ satisfies $p_{\rm diag}<\frac{3}{4}$,
the states $\varrho$ and $\xi$ are unambiguously different.

%%%%%%%%%%%%%%%%%%%%%%%%%%%%%%%%%%%%%%%%%%%%%%%%%%%%%%%%%%%%%%%%%%%
%%%%%%%%%%%%%%%%%%%%%%%%%%%%%%%%%%%%%%%%%%%%%%%%%%%%%%%%%%%%%%%%%%%
\begin{figure*}
\includegraphics{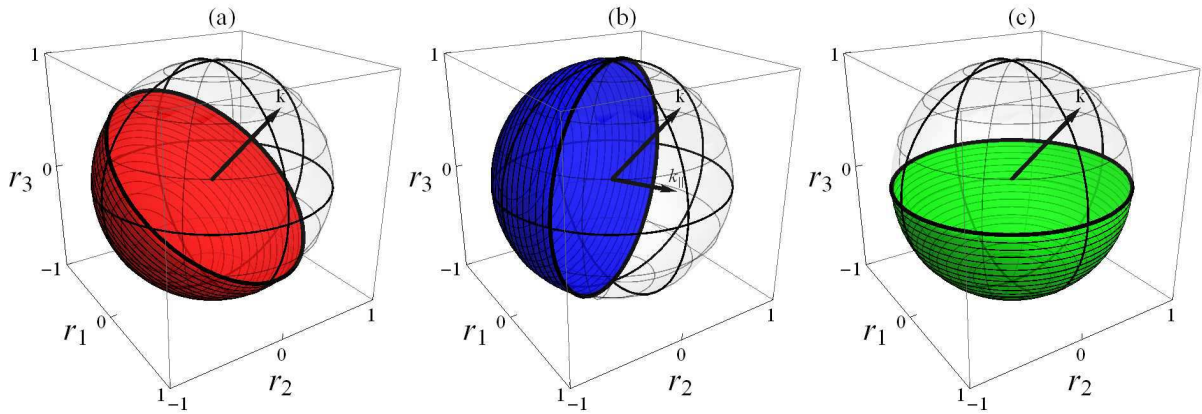}
\caption{\label{figure:fixed_pair_diagonal} (Color online) States
$\varrho$ in the Bloch ball (determined by vectors ${\bf r}$)
which can be distinguished from a fixed state $\xi$ (given by
vector ${\bf k}$) by using measurement $\E_{\rm SWAP}$ (a),
$\E_{xy}$ (b), and $\E_{z}$ (c).}
\end{figure*}
%%%%%%%%%%%%%%%%%%%%%%%%%%%%%%%%%%%%%%%%%%%%%%%%%%%%%%%%%%%%%%%%%%%
%%%%%%%%%%%%%%%%%%%%%%%%%%%%%%%%%%%%%%%%%%%%%%%%%%%%%%%%%%%%%%%%%%%

%-------------------------------------------------------------------------------------------

\subsubsection{Fixed pair comparison}

Surprisingly, there are pairs of states $\varrho$ and $\xi$ such
that no measurement $E$ of the form (\ref{E-diag}) can reveal
their difference. A direct calculation gives that the difference
for a pair of states $\varrho$ and $\xi$ can be concluded
in the diagonal comparison experiment if
\begin{equation} \label{RK_1} D_{{\rm diag}}(\varrho\otimes\xi,\cS^+)=\min_{|{\bf h}| \le 1}
\left| \textstyle\sum_{m=1}^3\varkappa_m (r_m k_m - h_m^2)\right|
> 0\,.
\end{equation}

\noindent Suppose that $r_m k_m$ is nonnegative for all $m$.
Setting $h_m=\sqrt{r_m k_m}$ the distance (\ref{RK_1}) is
vanishing for arbitrary measurement of the considered diagonal
form. Let us stress that the requirement of positivity of $r_m
k_m$ for all $m$ means that signs of the Bloch vector components
coincide, hence, ${\bf k}$ and ${\bf r}$ belong to the same octant
of the Bloch ball. Let us stress, however, that the octants depend
on the choice of the axes (Pauli operators), and for a given pair
of states we can always fix the coordinate system in such a way
that they belong to two different octants. The only exceptions are
collinear vectors ${\bf k}$ and ${\bf r}=c{\bf k}$ for $c\geq 0$.
In fact, a pair of parallel Bloch vectors (pointing in the same
direction) is indistinguishable by any diagonal measurement
irrelevant of the choice of coordinate system. In particular, it
follows that none of these measurements is capable of
distinguishing (in the comparison sense) the complete mixture
$\varrho=\frac{1}{2}I$ from any other state, because $p_{\rm
diff}(\frac{1}{2}I\otimes\xi)= p_{\rm
same}(\frac{1}{2}I\otimes\frac{1}{2}I)=\varkappa_0$.

It is natural to ask for which pairs $\varrho,\xi$ their
difference can be identified by a suitably selected $E$ of the
considered diagonal form and whether there are some
``non-diagonal" measurements enabling us to compare a pair of
states containing the complete mixture.

In order to get an insight into the power of diagonal
measurements, let us assume that $\varkappa_m\geq 0$, $m=1,2,3$
and fix ${\bf k}$. Define a new vector ${\bf K}=(\varkappa_1
k_1,\varkappa_2 k_2,\varkappa_3 k_3)$. If ${\bf K}\cdot{\bf r}>0$,
then we can find ${\bf h}$ such that ${\bf K}\cdot{\bf
r}=\sum_{m=1}^3 \varkappa_m h_m^2$, hence $D_{{\rm
diag}}(\varrho\otimes\xi,\cS^+)=0$. If ${\bf K}\cdot{\bf r}<0$,
then $D_{{\rm diag}}(\varrho\otimes\xi,\cS^+) = |{\bf K}\cdot{\bf
r}|+ \sum_{m=1}^3 \varkappa_m h_m^2 > 0$ for any ${\bf h}\neq {\bf
0}$. Therefore, the minimum is achieved for $\eta\otimes\eta
=\frac{1}{2}I\otimes\frac{1}{2}I$ and the distance reads
\begin{equation}
\label{RK_2} D_{{\rm diag}}(\varrho\otimes\xi,\cS^+)=
\left\{\begin{array}{cl}
0 & {\rm if}\ {\bf K}\cdot{\bf r}\geq 0,\\
|{\bf K}\cdot{\bf r}| & {\rm otherwise}.
\end{array}
\right.
\end{equation}

In other words, the considered diagonal measurement $\E$ enables
us to verify the difference between $\varrho$ and $\xi$ for all
$\varrho$ satisfying the inequality ${\bf K}\cdot{\bf r}<0$. The
condition ${\bf K}\cdot{\bf r}=0$ determines a plane containing
the complete mixture (center of the Bloch ball), hence, for any
measurement of the considered type and any state $\xi$ the set of
successfully comparable states $\varrho$ is exactly a hemisphere
of the Bloch ball. Fig.~\ref{figure:fixed_pair_diagonal}
illustrates this situation for the following choices of the
diagonal measurements (POVMs):
\begin{eqnarray}
\label{anti-} \E_{\rm SWAP} &=& \big\{ E_{\rm sym}=\tfrac{1}{4}
\big( 3\cdot I\otimes I + \textstyle\sum_{m=1}^3
\sigma_m\otimes\sigma_m \big), \quad \nonumber\\
&& E_{\rm asym}=\tfrac{1}{4} (I\otimes I -
\textstyle\sum_{m=1}^3 \sigma_m\otimes\sigma_m) \big\} ; \\
\label{Pxy} \E_{xy} &=& \big\{ E_{xy+} = \tfrac{1}{4} \big(3 \cdot
I\otimes I + \textstyle\sum_{m=1}^2
\sigma_m\otimes\sigma_m \big),\nonumber\\
&& E_{xy-}= \tfrac{1}{4} \big( I\otimes I-\textstyle\sum_{m=1}^2
\sigma_m\otimes\sigma_m \big) \big\};\\
\label{Pz} \E_z &=& \big\{ E_{z\pm}=\tfrac{1}{2} (I\otimes I \pm
\sigma_3\otimes\sigma_3) \big\}.
\end{eqnarray}

In particular, for $\E_{\rm SWAP}$ the ``comparable hemisphere" is
orthogonal to the vector ${\bf k}$. For $\E_{xy}$ the
``comparable" hemisphere is orthogonal to the vector ${\bf
k}_{\parallel}=(k_1,k_2,0)$ being a projection of ${\bf k}$ onto
the $xy$ plane. Finally, for $\E_z$ any state from the northern
hemisphere is ``comparable" with any state from the southern
hemisphere.

It is worth noting that we have restricted ourselves to the
specific form of POVM effects (\ref{E-diag}). However, even for
such a simplified problem the solution looks rather sophisticated.

%-------------------------------------------------------------------------------------------

\subsubsection{Average performance}
The fact that for any given diagonal two-valued measurement the
states within the same octant are not comparable means that none
of them is universal neither in an approximative way.
Nevertheless, it is of interest to understand which of them
perform better than the others and which do not perform at all. In
particular, we are interested in the answer to the following
question: How many qubit states $\varrho,\xi$ can be
distinguished? As is briefly outlined in
Sec.~\ref{section:universal}, to answer this question it is
necessary to introduce some measure $\mu$ on the state space. Once
it is introduced, we can evaluate the quantities $|S^-_{\rm
comp}|_\E$ (relative volume of the successfully comparable states,
i.e. the comparison universality factor) and $\langle D_\E\rangle$
(average distance, i.e. the comparison quality factor).

In contrast with pure states, for density operators there exist
many equivalently well-motivated measures (see, e.g.,
\cite{Hall,Bengtsson} and references therein). We will employ two
most commonly used ones, namely, the Hilbert--Schmidt measure
$\mu_{\rm HS}$ and the Bures measure $\mu_{\rm B}$:
\begin{eqnarray}
\label{HS} \mu_{\rm HS}(d\varrho) &=& \frac{3}{4\pi} \, r^2 \sin\theta \, d r \, d\theta \, d\varphi \, ,\\
\label{bures} \mu_{\rm B}(d\varrho) &=&  \frac{r^2 \sin\theta
}{\pi^2 \sqrt{1-r^2}} \, d r \, d\theta \, d\varphi \, ,
\end{eqnarray}

\noindent where we used the following parametrization of Bloch
vectors: ${\bf r} = (r\cos\varphi\sin\theta,
r\sin\varphi\sin\theta, r\cos\theta)$ with $r\in[0,1]$,
$\theta\in[0,\pi]$, and $\varphi\in[0,2\pi]$. Both these measures
are spherically symmetric and the former one corresponds to the
uniform coverage of the entire Bloch ball~\cite{Hall}, i.e. the
Hilbert--Schmidt measure $\mu_{\rm HS}(T)$ of any compact set $T
\subset \cS(\cH_2)$ equals the geometrical volume
$\int_{\varrho({\bf r}) \in T} d^3 {\bf r}$ of the corresponding
body inside the Bloch ball divided by $4\pi/3$. The Bures measure
(\ref{bures}) ascribes higher weights to the states with higher
purity (that are closer to the surface of the Bloch ball).

%%%%%%%%%%%%%%%%%%%%%%%%%%%%%%%%%%%%%%%%%%%%%%%%%%%%%%%%%%%%%%%%%%%
%%%%%%%%%%%%%%%%%%%%%%%%%%%%%%%%%%%%%%%%%%%%%%%%%%%%%%%%%%%%%%%%%%%
\begin{figure*}
\includegraphics[width=18cm]{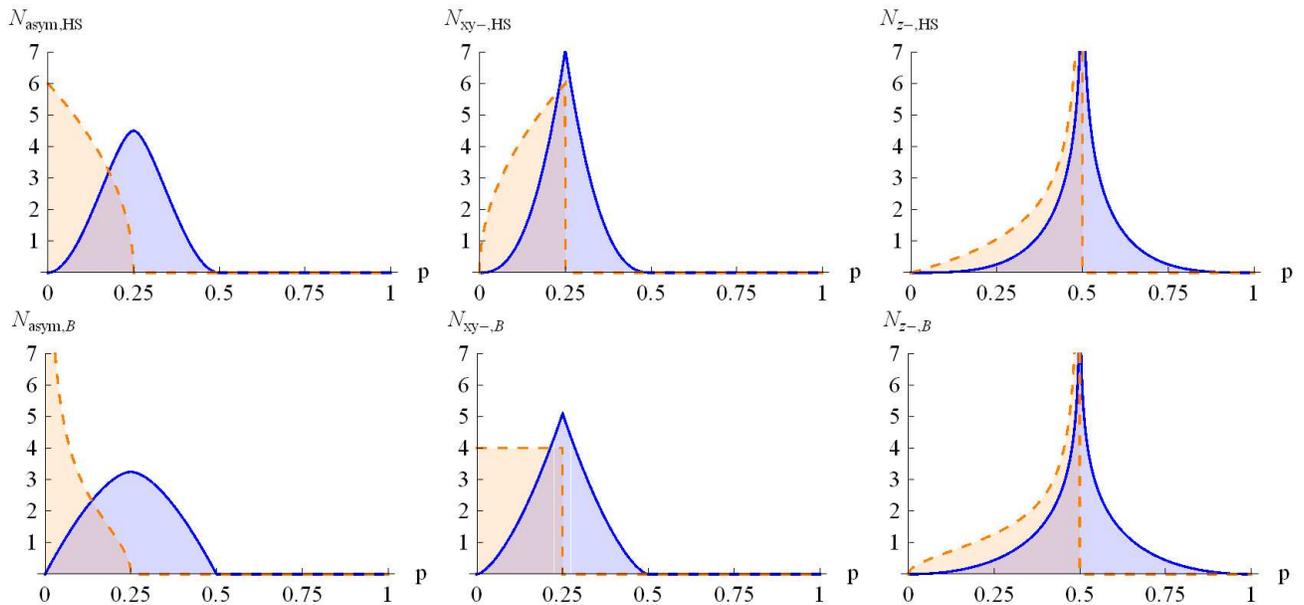}
\caption{\label{figure:measures} (Color online) Densities of
states $N_{E,\mu}^{\rm diff}(p)$ (solid) and $N_{E,\mu}^{\rm
same}(p)$ (dashed) for different POVM effects: $E_{\rm asym}$
(left column), $E_{xy-}$ (middle column), and $E_{z-}$ (right
column); and for different measures: the Hilbert--Schmidt measure
(top row) and the Bures measure (bottom row).}
\end{figure*}
%%%%%%%%%%%%%%%%%%%%%%%%%%%%%%%%%%%%%%%%%%%%%%%%%%%%%%%%%%%%%%%%%%%
%%%%%%%%%%%%%%%%%%%%%%%%%%%%%%%%%%%%%%%%%%%%%%%%%%%%%%%%%%%%%%%%%%%

For calculation purposes it is convenient to introduce the
following (relative) density of states:
\begin{equation}
\label{density-diff} N_{E,\mu}^{\rm diff}(p) = \lim_{\Delta p
\rightarrow 0} \frac{1}{\Delta p} \iint_{ \tr{E \varrho\otimes\xi}
\in [p;p + \Delta p]} \mu(d\varrho)\mu(d\xi)\, ,
\end{equation}

\noindent whose physical meaning is that $N_{E,\mu}^{\rm diff}(p)
\Delta p$ equals the fraction of pairs $\varrho\otimes\xi$
resulting in the measurement outcome probability within the region
$[p,p+\Delta p]$ for the effect $E$. Using the introduced
function, we can readily write
\begin{eqnarray}
&& |S^-_{\rm comp}|_{\E} = \int_{p \in P_{E}^-\setminus P_{E}^+}
N_{E,\mu}^{\rm diff}(p) d p\,,\\
&& \label{Dist-mean} \langle D_\E\rangle = 2 \int_{p\in
P_{E}^-\setminus P_{E}^+} \left|p-p_0\right| N_{E,\mu}^{\rm
diff}(p) dp\,, \qquad\qquad
\end{eqnarray}

\noindent where $P_{E}^{\pm}$ stands for the image of ${\cal
S}^{\pm}$ under the action of POVM effect $E$, $p_0$ is
simultaneously the frontier point of $P_{E}^+$ and the inner point
of $P_{E}^-$ (if there are two such points, then $\langle D_\E
\rangle$ is a sum of two integrals (\ref{Dist-mean}) in the
corresponding regions of variable $p$).

In what follows we will analyze the three examples from the
previous section (comparison measurements $\E_{\rm SWAP}$,
$\E_{xy}$, and $\E_{z}$) and compare their performance. The
associated densities are depicted in Fig.~\ref{figure:measures}
and explicitly written in the Appendix. We focus on these POVMs
because they represent three different types of boundary (extremal
in case of $\E_{\rm SWAP}$ and $\E_{z}$) points of the set of
diagonal measurements (see Fig.~\ref{figure:Rhombo}). Other
diagonal measurements will exhibit intermediate behavior with
respect to these three.

We have numerically tested that for diagonal measurements the
relative volume of the set of comparable states $|\cS_{\rm
comp}^+|_\E \leq \frac{1}{2}$ irrelevant of the measure used. The
considered three examples saturate this value, i.e. $|\cS_{\rm
comp}^+|_{\E_{{\rm SWAP},xy,z}} = \frac{1}{2}$. In fact, there
exist many measurements (of the considered family) for which the
comparable set is of this size and the question is whether there
are some interesting differences in their performance. As a figure
of merit for this purposes we employ the average distance $\langle
D_\E\rangle$, which is closely related to the quality of the
fixed-pair comparison and partially quantifies also the difference
of states.

The performance of three comparison measurements $\E_{\rm SWAP}$,
$\E_{xy}$, and $\E_{z}$ is compared in
Table~\ref{table:comparison}. It is clear that whichever measure
$\mu$ is used, the mean value $\langle D_\E \rangle_\mu$ is
greater for the measurements $\E_{\rm SWAP}$ and $\E_{z}$ than
that for the measurement $\E_{xy}$. Furthermore, although both
POVMs $\E_{\rm SWAP}$ and $\E_{z}$ lead to the same expectation
values $\ave{D_{\E}}_{\mu}$, the former one gives rise to less
dispersion $\mathsf{D}_{\mu}\left[D_{\E}\right]$ and relative
standard deviation $\sqrt{\mathsf{D}_{\mu}\left[D_{\E}\right]} /
\ave{D_{\E}}_{\mu}$. In addition, from Fig.~\ref{figure:measures}
it follows that the effect $E_{\rm asym}$ results in the smallest
density of states in the vicinity of the point $p_0$. Such a
feature is very demanding because the values close to $p_0$ are
the most affected by potential statistical errors, which are
unavoidable in practice.

\begin{table}
\caption{\label{table:comparison} Effectiveness of
probability-based comparison based on different POVM effects. Mean
values $\ave{\cdot}_{\mu}$ and dispersions $\mathsf{D}_{\mu}
[\cdot]$ of the distance $D_{\E} (\varrho\otimes\xi,\cS^+)$ for
different kinds of measure $\mu$.}
\begin{ruledtabular}
\begin{tabular}{cccc}
 & $\E_{\rm SWAP}$ & $\E_{xy}$ & $\E_{z}$\\
\hline $\ave{D_{\E}}_{\rm HS}$ & 0.07032 & 0.05522 & 0.07032 \\
$\ave{D_{\E}}_{\rm B}$ & 0.09006 & 0.07074 & 0.09006 \\
$\mathsf{D}_{{\rm HS}}\left[D_{\E}\right]$
& 0.01006 & 0.00695 & 0.01506 \\
$\mathsf{D}_{{\rm B}}\left[D_{\E}\right]$
& 0.01533  & 0.01062 & 0.02314 \\
$\sqrt{\mathsf{D}_{{\rm HS}}\left[D_{\E}\right]} /
\ave{D_{\E}}_{\rm HS}$
& 1.426 & 1.510 & 1.745 \\
$\sqrt{\mathsf{D}_{{\rm B}}\left[D_{\E}\right]} /
\ave{D_{\E}}_{\rm B}$
& 1.375 & 1.457 & 1.689 \\
\end{tabular}
\end{ruledtabular}
\end{table}

Using the mentioned figures of merit, we can draw a conclusion
that the measurement $\E_{\rm SWAP}$ performs (on average) the
best among the considered measurements. There is yet another fact
in favor of this. Fig.~\ref{figure:measures} contains also the
densities of the same states  $N_{E,\mu}^{\rm same}(p)$ defined by
\begin{equation}
\label{density-same} N_{E,\mu}^{\rm same}(p) = \lim_{\Delta p
\rightarrow 0}\frac{1}{\Delta p} \int_{\tr{E \eta\otimes\eta}\in
[p;p + \Delta p]} \mu(d\eta).
\end{equation}

\noindent The value of the quantity $N_{E,\mu}^{\rm same}(p)\Delta
p$  tells us the number of states of the form $\eta\otimes \eta$
for which the probability ${\rm tr}[E \eta\otimes\eta]$ belongs to
the region $[p;p + \Delta p]$ (explicit formulas for the involved
effects $E$ are given in the Appendix). One can clearly see from
Fig.~\ref{figure:measures} that for the SWAP-based comparison
(unlike the other two) the distribution is concentrated far from
the border point $p_0$. It is evident that if the measured
experimentally probability $p$ satisfies $p \in P_{\rm asym}^+$
then one cannot judge whether states are the same or different.
However, even in this case it is possible to extract additional
information. In fact, the measured probability $p$ sets a limit on
the maximum trace distance between the states $\varrho$ and $\xi$
(provided they are different), because ${\rm
tr}{|\varrho-\xi|}\leq 2\sqrt{2p}$. Consequently, the smaller the
measured probability $p$ the closer the states $\varrho$ and $\xi$
are.

%-------------------------------------------------------------------------------------------

\subsection{``Non-diagonal" qubit measurements lacking in almost universality}
We have argued that ``diagonal" qubit measurements are not able to
decide on the difference of states defined by codirectional Bloch
vectors, in particular, the states $\frac{1}{2}I\otimes\varrho$
are not in the comparable sets of any measurement from this
family. We address the question whether this feature is general.
In other words, whether there exists a two-valued qubit
measurement allowing us to decide on the difference between the
complete mixture and some other state.

%%%%%%%%%%%%%%%%%%%%%%%%%%%%%%%%%%%%%%%%%%%%%%%%%%%%%%%%%%%%%%%%%%%
%%%%%%%%%%%%%%%%%%%%%%%%%%%%%%%%%%%%%%%%%%%%%%%%%%%%%%%%%%%%%%%%%%%
\begin{figure*}
\includegraphics{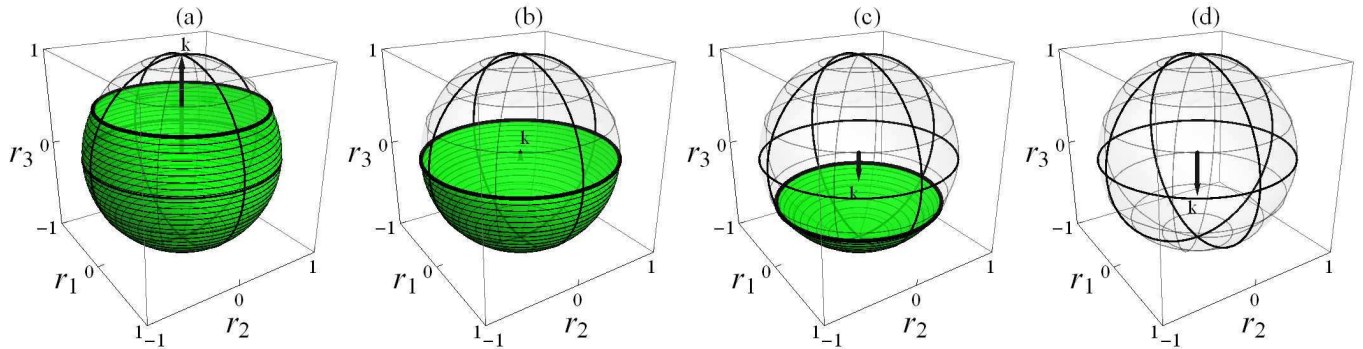}
\caption{\label{more-than-half} (Color online) States $\varrho$
inside the Bloch ball (determined by vectors ${\bf r}$) which can
be distinguished from a particular fixed state $\xi = {\rm
diag}(\Xi_1,\Xi_2)$ (given by vector ${\bf k}=(0,0,\Xi_1 -
\Xi_2)$) by using the non-diagonal POVM effect $E =
\ket{\Xi_2\otimes\Xi_1}\bra{\Xi_2\otimes\Xi_1} \equiv {\rm
diag}(0,0,1,0)$ for the following cases: $\xi={\rm diag}(1,0)$
(a), $\xi={\rm diag}(\frac{1}{2},\frac{1}{2})$ (b), $\xi={\rm
diag}(\frac{1}{3},\frac{2}{3})$ (c), and $\xi={\rm
diag}(\frac{1}{4},\frac{3}{4})$ (d).}
\end{figure*}
%%%%%%%%%%%%%%%%%%%%%%%%%%%%%%%%%%%%%%%%%%%%%%%%%%%%%%%%%%%%%%%%%%%
%%%%%%%%%%%%%%%%%%%%%%%%%%%%%%%%%%%%%%%%%%%%%%%%%%%%%%%%%%%%%%%%%%%

Consider a qubit state $\xi$ and its spectral decomposition $\xi=
\sum_{i=1,2} \Xi_i \ket{\Xi_i}\bra{\Xi_i}$ with eigenvalues
$\Xi_1\geq\Xi_2$. Suppose a two-valued measurement $\E$ with the
effects $E_1=\ket{\Xi_2\otimes\Xi_1}\bra{\Xi_2\otimes\Xi_1}$ and
$E_2=I-E_1$. Then we have
\begin{eqnarray}
D_{\E}(\varrho\otimes\xi,\mathcal{S}^{+}) &=& 2
\inf_{\eta\otimes\eta \in \mathcal{S}^{+}}
|\varrho_{22}\Xi_1 - \eta_{11}\eta_{22}| \nonumber\\
&=& \left\{
\begin{array}{cc}
  2 \varrho_{22}\Xi_1 - \frac{1}{2}  & {\rm if}~ \varrho_{22}>(4\Xi_1)^{-1}, \\
  0 & {\rm otherwise},\\
\end{array}\right. \qquad
\end{eqnarray}

\noindent where $\varrho_{ii} = \bra{\Xi_i} \varrho \ket{\Xi_i}$
and $\eta_{ii} = \bra{\Xi_i} \eta \ket{\Xi_i}$, $i=1,2$. If, for
instance, $\Xi_1=1$ (i.e. $\xi$ is the north pole of the Bloch
ball), then states $\varrho$ with $\varrho_{22}>\frac{1}{4}$ form
the comparable set for $\xi$. They lie below latitude 60$^{\circ}$
North and include also the maximally mixed state (see
Fig.~\ref{more-than-half}a). If $\xi$ is the maximally mixed state
($\Xi_1=\Xi_2=\frac{1}{2}$), then it is unambiguously
distinguished from any state from the southern hemisphere of the
Bloch ball (Fig.~\ref{more-than-half}b). Applying unitary
transformations of the form $U\varrho U^{\dag}$ and $(U\otimes
U)E_{1,2}(U^{\dag}\otimes U^{\dag})$, we can draw a conclusion
that maximally mixed state can be effectively compared with any
other qubit state, which answers our question in a positive way.

We find that on average the fraction of comparable states is
smaller than $\frac{1}{2}$. In particular, $|\cS_{\rm
comp}^-|_{\rm HS}=\frac{3}{8}(6-7\ln 2)=0.43$ and $|\cS_{\rm
comp}^-|_{\rm B}=0.42$, which means that the measurements $\E_{\rm
SWAP}$, $\E_{xy}$, and $\E_{z}$ outperform the considered
non-diagonal measurement in this parameter. The average distance
$\ave{D_{\E}}$ reads 0.1342 and 0.1524, respectively. We can see
that the quality factor $\ave{D_{\E}}$ is increased by the expense
of the lower universality factor $|\cS_{\rm comp}^-|$.

Surprisingly, for a fixed pure state $\xi = \ket{\Xi}\bra{\Xi}$ it
is possible to design a two-outcome measurement $\E_{\Xi}$ such
that $D_{{\Xi}}(\varrho\otimes\xi,\cS^+) > 0$ for all states
$\varrho \ne \xi$. In fact, suppose a POVM $\E_{\Xi}$ with effects
$E_{\Xi 1} = {\rm
diag}(\frac{1}{4},\frac{3}{8},\frac{1}{8},\frac{5}{8})$ and
$E_{\Xi 2} = I-E_{\Xi 1}$ specified in the orthonormal basis $\{
\ket{\Xi\otimes\Xi}, \ket{\Xi\otimes\Xi_\perp},
\ket{\Xi_\perp\otimes\Xi}, \ket{\Xi_\perp\otimes\Xi_\perp} \}$.
Then, $p_{{\Xi 1}}(\varrho\otimes\xi)= \frac{1}{8} (2-
\varrho_{22})$ and $p_{{\Xi 1}}(\eta\otimes\eta)=
\frac{1}{8}(2+3\eta_{22}^2)$, where
$\varrho_{22}=\langle\Xi_{\perp}|\varrho|\Xi_{\perp}\rangle$ and
$\eta_{22}=\langle\Xi_{\perp}|\eta|\Xi_{\perp}\rangle$. Hence
$p_{{\Xi 1}}(\varrho\otimes\xi) \in [\frac{1}{8},\frac{1}{4})$ and
$p_{{\Xi 1}}(\eta\otimes\eta) = [\frac{1}{4},\frac{5}{8}] \equiv
P_{{\Xi 1}}^+$. Therefore, if $\varrho\neq\xi$, then we
necessarily observe a probability $p_{{\Xi 1}}$ outside the
interval $P_{{\Xi 1}}^+$, which unambiguously identifies the
difference of $\varrho$ and $\xi$. Such a two-outcome experiment
can be used to check whether a copy $\varrho$ of the etalon pure
state $\xi$ was really produced. Nonetheless, in spite of the
seeming effectiveness of this measurement, its average performance
is quite low. To be precise, the fraction of comparable states
$\varrho\otimes\xi$ on average reads $|\cS_{\rm comp}^-|_{\rm
HS}=0.097$, or $|\cS_{\rm comp}^-|_{\rm B}=0.131$, and the quality
factor is $\ave{D_{\Xi}}_{\rm HS} = 0.0049$, or
$\ave{D_{\Xi}}_{\rm B} = 0.0079$.

%-------------------------------------------------------------------------------------------

\section{\label{section:AUC-measurement} Almost universal comparison measurements}

We have shown in Sec.~\ref{section:universal} that any universal
comparison measurement is necessarily an LIC measurement. However,
the question of an existence of an almost universal comparison
measurement (which is not LIC) remains open. In
Sec.~\ref{section:two-valued}, we have considered such examples of
two-valued measurements that the fraction of unambiguously
comparable qubit states $\varrho$, $\xi$ does not exceed
$\frac{1}{2}$. In what follows we find a two-valued POVM, for
which this fraction equals $1$. This makes such a two-outcome
measurement almost universal.

Consider a general two-valued measurement $\E_{\text{2-out}}$
composed of the effects $E={\rm
diag}(\lambda_1,\lambda_2,\lambda_3,\lambda_4)$ and $I-E$ that are
diagonal in some Hilbert space basis $\ket{j\otimes k}$ of two
qubits. For $\E_{\text{2-out}}$ to be almost universal, the set of
indistinguishable states $\varrho\otimes\xi$ ($\varrho\ne\xi$)
must have measure zero. Let us recall the geometrical picture
presented in the beginning of Sec.~\ref{section:two-valued}: for
any fixed qubit state $\xi$ the comparable states $\varrho$ form a
cut or two cuts of the Bloch ball, in latter case the cuts being
separated by the distance $L$ (see formula (\ref{L-distance})).
The almost universality requires $L=0$ for all states $\xi$ from a
set of measure 1 in $\cS(\cH_2)$. In other words, the almost
universality requires $\max_{\eta\otimes\eta\in\cS^{+}}
\tr{E\eta\otimes\eta} = \min_{\eta\otimes\eta\in\cS^{+}}
\tr{E\eta\otimes\eta}$, i.e. $\tr{E\eta\otimes\eta} = {\rm const}$
for all $\eta\in\cS(\cH)$. On the other hand, the only invariant,
which is quadratic with respect to $\eta$, is $(\tr{\eta})^2$,
which means that $\tr{E\eta\otimes\eta} = \lambda_1 \eta_{11}^2 +
(\lambda_2+\lambda_3) \eta_{11}\eta_{22} + \lambda_4 \eta_{22}^2
\propto \eta_{11}^2 + 2 \eta_{11}\eta_{22} + \eta_{22}^2$, i.e.
$\lambda_1=\lambda_4=\frac{1}{2}(\lambda_2+\lambda_3) \equiv
\lambda \in[0,1]$. Denoting $\frac{1}{2}(\lambda_2-\lambda_3)
\equiv \mu \in[-\min(\lambda,1-\lambda),\min(\lambda,1-\lambda)]$,
the operator $E={\rm
diag}(\lambda,\lambda+\mu,\lambda-\mu,\lambda)$ is an effect
indeed. The constructed effect determines an almost universal
comparison because $P_{E}^+ = \{\lambda\}$ and $P_{E}^- =
[\lambda-|\mu|,\lambda+|\mu|]$. The distance
(\ref{distance-two-valued}) is easily calculated and reads
\begin{equation}
\label{distance-2-au} D_{\text{2-out}} (\varrho \otimes \xi,
\cS^+) = 2 |\mu (\varrho_{11}-\xi_{11})|.
\end{equation}

\begin{figure}
\includegraphics[width=8cm]{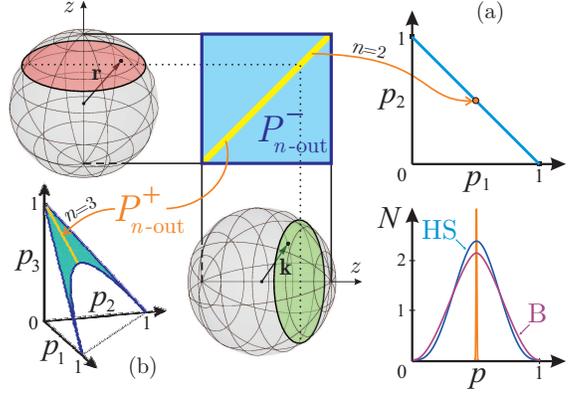}
\caption{\label{fig:almost_universal} (Color online) Almost
universal comparison measurement (for qubits) with $n$ outcomes:
(a) $n=2$, the images $P_{\text{2-out}}^{\pm}$ on 1-simplex and
the densities $N_{E,\mu}^{\rm diff}(p)$ for Hilbert--Schmidt (HS)
and Bures (B) measures, with $\delta$-peak being ascribed to
$N_{E,\mu}^{\rm same}(p)$; (b) $n=3$, the images
$P_{\text{3-out}}^{\pm}$ on 2-simplex. In both cases the states
$\varrho$ and $\xi$ can be distinguished whenever their Bloch
vectors satisfy $r_z \ne k_z$.}
\end{figure}

Let the basis $\ket{j\otimes k}$ be composed of eigenvectors of
$\sigma_z\otimes\sigma_z$. For a given state $\xi$ the set of
comparable states $\varrho$ equals the whole Bloch ball except for
a circle of states satisfying $r_z=k_z$, where $r_z$ and $k_z$ are
$z$-components of the corresponding Bloch vectors ${\bf r}$ and
${\bf k}$, respectively (see Fig.~\ref{fig:almost_universal}a).

The distance (\ref{distance-2-au}) takes maximal value for maximal
$\mu$, i.e. when $\mu = \lambda = \frac{1}{2}$. In this case
$E={\rm diag}(\frac{1}{2},1,0,\frac{1}{2})$. The quality factor of
such two-valued almost universal comparison is
$\ave{D_{\text{2-out}}}_{\rm HS} = 18/35 \approx 0.51$ or
$\ave{D_{\text{2-out}}}_{\rm B} = 256/45\pi^2 \approx 0.58$, i.e.
substantially greater than for ``diagonal'' two-valued
measurements (cf. Sec.~\ref{section:two-valued}).  The densities
(\ref{density-diff}) and (\ref{density-same}) for the effect $E$
are given in the Appendix and depicted also in
Fig.~\ref{fig:almost_universal}a.

The qubit example of two-valued almost universal comparison
measurement can be straightforwardly generalized to any dimension
$d$. In fact, the choice $E=\frac{1}{2}(A\otimes I +
I\otimes(I-A))$, where $O \le A \le I$, guarantees that
$\tr{E\eta\otimes\eta} = \frac{1}{2}$ for all states
$\eta\in\cS(\cH)$. Therefore the distance (\ref{L-distance})
vanishes and the comparison is almost universal. The distance
(\ref{distance-two-valued}) equals $D_{\text{2-out}} (\varrho
\otimes \xi, \cS^+) = \left| \tr{A(\varrho-\xi)} \right|$. For
conventionally parameterized states $\varrho=\frac{1}{d}(I+{\bf
r}\cdot\boldsymbol{\Lambda})$ and $\xi=\frac{1}{d}(I+{\bf
k}\cdot\boldsymbol{\Lambda})$, one can choose $A = A_j \equiv
 \frac{1}{2}(I + \sqrt{\frac{2}{d}} \Lambda_j)$ that ensures $O \le A_j \le I$. Then $D_{\text{2-out}_j} (\varrho
\otimes \xi, \cS^+) = \frac{1}{\sqrt{2d}} |r_j - k_j| > 0$
whenever $r_j \ne k_j$.

The considered two-valued almost universal measurement
$\E_{\text{2-out}_j}$ compares $j$-components of
$(d^2-1)$-dimensional Bloch-like vectors ${\bf r}$ and ${\bf k}$.
Combining all these measurements $\E_{\text{2-out}_j}$,
$j=1,\ldots,d^2-1$, into a single measurement allows us to
distinguish vectors ${\bf r}$ and ${\bf k}$, i.e. performing
universal comparison measurements with $2(d^2-1)$ effects. Such a
measurement will be LIC, in total agreement with the results of
Sec.~\ref{section:universal}.

Although a measurement with two outcomes can be almost universal,
the practical realization of two-outcome measurements can be
rather complicated. In some physical experiments, many-valued
measurements are naturally performed instead of two-valued ones.
In view of this fact, it is reasonable to consider almost
universal many-valued measurements, which are closer to practical
realization.

We begin with the qubit case and three-valued measurement
$\E_{\text{3-out}}$ composed of effects $E_1={\rm diag}(0,1,0,0)$,
$E_2={\rm diag}(0,0,1,0)$, and $E_3={\rm diag}(1,0,0,1)$ defined
in the basis $\ket{j\otimes k}$. For twin-identical states
$\eta\otimes\eta$ the probabilities of outcomes satisfy $p_1^{\rm
same}=p_2^{\rm same} \le \frac{1}{4}$, $p_3^{\rm same}=1-p_1^{\rm
same}-p_2^{\rm same}\ge \frac{1}{2}$. In other words, the set
$\cS^+$ is mapped onto a line inside the probability simplex (see
Fig.~\ref{fig:almost_universal}b). On the other hand, the elements
of $\cS^-$, $\varrho\otimes\xi$, give rise to probability vectors
$\vec{p}^{\rm ~diff}$ intersecting the line of twin-identical
states, $P_{\text{3-out}}^+$, if and only if density matrices
$\varrho$ and $\xi$ have the same diagonal elements. However, the
subset of the states $\varrho\otimes\xi$ satisfying this peculiar
requirement has zero measure in $\cS^{-}$. Hence, almost all pairs
of states $\varrho$ and $\xi$ can be compared by the described
three-outcome measurement $\E_{\text{3-out}}$
(Fig.~\ref{fig:almost_universal}). A direct calculation of the
distance (\ref{eq:dist_1}) yields
\begin{equation}
D_{\text{3-out}} (\varrho \otimes \xi, \cS^+) = \frac{1}{2}
\left\{
\begin{array}{cc}
  |r_z - k_z| & {\rm if} ~ r_z  k_z \ge 0, \\
  |r_z - k_z| + |r_z  k_z| & {\rm otherwise}. \\
\end{array}
\right.
\end{equation}

\noindent It is not hard to see that the calculation of $|\cS_{\rm
comp}^-|$ by formula (\ref{S-comp}) results in $1$, i.e. the
comparison measurement $\E_{\text{3-out}}$ is indeed almost
universal. The average distance (\ref{D-E-ave}) is
$\ave{D_{\text{3-out}}}_{\rm HS} = 0.29$, or
$\ave{D_{\text{3-out}}}_{\rm B} = 0.33$. The quality factor is
smaller than that for two-valued almost universal measurement
because the average probabilities of outcomes are smaller
($\sim\frac{1}{3}$ vs. $\sim\frac{1}{2}$ for two-valued
measurements).

%%%%%%%%%%%%%%%%%%%%%%%%%%%%%%%%%%%%%%%%%%%%%%%%%%%%%%%%%%%%%%%%%%%
%%%%%%%%%%%%%%%%%%%%%%%%%%%%%%%%%%%%%%%%%%%%%%%%%%%%%%%%%%%%%%%%%%%
%%%%%%%%%%%%%%%%%%%%%%%%%%%%%%%%%%%%%%%%%%%%%%%%%%%%%%%%%%%%%%%%%%%
%%%%%%%%%%%%%%%%%%%%%%%%%%%%%%%%%%%%%%%%%%%%%%%%%%%%%%%%%%%%%%%%%%%

Let us note that the considered example of a 3-outcome POVM is
nothing else but a coarse-graining of a local projective
measurement applied on each of the system independently. In
particular, $E_1=E_{01}$, $E_2=E_{10}$, and $E_3=E_{00}+E_{11}$,
where $E_{jk}=\ket{j\otimes k}\bra{j\otimes k}$ ($j,k=0,1$) are
the effects forming the local (factorized) projective measurement
$\E_{\text{4-out}}$. In other words, if one performs the same
(along the same direction) Stern-Gerlach experiment on both
spin-$\frac{1}{2}$ systems, then the resulting four-outcome POVM
$\E_{\text{4-out}}$ performs an almost universal comparison. So
does the POVM $\E_{\text{3-out}}$, where the outcomes `00' and
`11' are unified into a single one.

This qubit example of three-outcome measurement can also be
generalized for an almost universal comparison of $d$-dimensional
systems performed by a measurement with $d(d-1)+1$ outcomes. Let
$\mathsf{Q}$ be a projective measurement associated with effects
$Q_j=\ket{\psi_j}\bra{\psi_j}$, where $\{\ket{\psi_j}\}_{j=1}^d$
form an orthonormal basis in ${\cal H}$. Suppose the same
measurement is performed on both $d$-dimensional systems and
define a coarse-grained POVM $\E_{\rm cg}$ with effects
$E_0=\sum_{j=1}^d Q_j\otimes Q_j$ and $E_{jk}=Q_j\otimes Q_k$ if
$j\neq k$. If the states are the same, then $p_{jk}^{\rm
same}=p_{kj}^{\rm same}$ and $p_0^{\rm same}=
1-2\sum_{j<k}p_{jk}^{\rm same}$. Clearly, $\vec{p}^{\rm ~diff}\in
P_{\rm cg}^+$ if $\bra{\psi_j}\varrho\ket{\psi_j} =
\bra{\psi_j}\xi\ket{\psi_j}$ for all $j=1,\ldots,d$. Thus, a
collection of all non-identical states $\varrho\otimes\xi \in
\cS^+$, whose images $\vec{p}^{\rm ~diff}\in P_{\rm cg}^+$, is a
subset of $\cS^-$ defined by $d-1$ real parameters (in view of
normalization) and, hence, this subset is less parametric than the
set $\cS^-$ defined by $2(d^2-1)$ real parameters, i.e. the
measure of this subset is zero. In conclusion, the measurement
$\E_{\rm cg}$ consisting of $d(d-1)+1$ effects $\{E_0, E_{jk}\}$
is an example of the almost universal comparison measurement for
$d$-dimensional quantum systems.

%-------------------------------------------------------------------------------------------

\section{\label{section:summary} Summary}
In its essence the comparison is a binary decision problem, which
we believe plays a very important role in our everyday lives. In
this paper we addressed its quantum version, namely, a comparison
of a pair of unknown sources of generally mixed states. We
designed a new comparison strategy based on the observed
statistics (not individual outcomes) of a particular comparison
measurement device. It turns out that, basing upon the observed
probabilities of measurement outcomes, one can sometimes draw an
unambiguous conclusion on the difference between the states. In
fact, it seems that a vast majority of measurements are capable to
compare some pairs of states. However, as we have shown in this
paper, the universal comparison of two arbitrary states requires
locally informationally complete measurements, hence, the complete
tomography of both sources is necessary and sufficient in order to
perfectly distinguish between twin-identical states
$\eta\otimes\eta$ and non-identical states $\varrho\otimes\xi$
($\varrho \ne \xi$).

Furthermore, we analyzed the comparison performance of two-valued
qubit measurements. We defined the family of ``diagonal"
measurements including the so-called SWAP-based comparison
measurement, which is known to be useful for the single-shot
unambiguous comparison of pure states. We have shown that none of
these diagonal measurements is able to decide on the difference of
a completely mixed state from any other mixed state. Consequently,
the fraction of comparable states, $|\cS^-_{\rm comp}|_\E$, is at
most $\frac{1}{2}$ for diagonal measurements. We compared in
detail the average performance of three diagonal measurements
$\E_{\rm SWAP}, \E_{xy}$, and $\E_z$ that are boundary (extremal
in case of $\E_{\rm SWAP}$ and $\E_z$) points of diagonal
measurements. Although for all of these measurements $|\cS^-_{\rm
comp}|_\E=\frac{1}{2}$, we found differences in the distribution
of distances $D_\E(\varrho\otimes\xi,\cS^+)$. Using these
considerations, we concluded that the SWAP-based comparison
performs (on average) better than the other two examples.

We also provided non-diagonal comparison measurements enabling us
to decide on the difference between an arbitrary state $\xi$ and
the complete mixture $\frac{1}{2}I$ or between a pure state $\xi$
and any other state (in both cases the measurement depends on
$\xi$). In this sense, for a given $\xi$ the measurements of this
kind overcome the performance of any diagonal measurement.
However, their average performance over the set of all states
results in $|\cS^-_{\rm comp}|_\E < \frac{1}{2}$. Despite this
shortcoming, any pair of qubit states can be compared in a
suitable non-diagonal two-valued measurement.

In the remaining part we presented the almost universal comparison
measurements in any dimension, i.e. such measurements that the
size of the comparable set is maximal, $|\cS^-_{\rm comp}|_\E=1$,
but there are still some pairs of states (forming a subset of
measure zero) for which their difference cannot be certified. The
almost universal comparison can even be realized by two-outcome
measurements. For qubits, the constructed two-valued almost
universal measurement is shown to exhibit the best average
performance. We succeeded in finding the explicit form of
two-valued almost universal comparators in any dimension.
Nonetheless, many-outcome almost universal comparison measurements
may turn out to be more feasible for practical implementation than
two-valued ones. For $d$-dimensional systems we theoretically
constructed such measurements with $d(d-1)+1$ outcomes ($3$ in
case of qubits). Each measurement is just a coarse-graining of a
local measurement, where both systems are measured by the same
($d$-valued) projective measurement (e.g., the Stern-Gerlach
apparatus oriented along $z$-direction in the case of spin
particles).

In summary, we have shown that the universal comparison of states
(in the considered settings) is not possible, but that there still
exist simple almost universal comparators. In particular,
two-outcome measurements are already sufficient (in any dimension)
for almost universality. A nice feature of the proposed
many-outcome almost universal comparison measurements is their
experimental simplicity. We left many open questions, especially
concerning an optimality of the almost universal comparison
measurements. As concerns the universal comparison, an optimality
is directly related to the optimal complete tomography.

%-------------------------------------------------------------------------------------------

\begin{acknowledgments}
The authors appreciate fruitful discussions with Tom\'{a}\v{s}
Ryb\'{a}r and thank the anonymous referee for insightful and
constructive comments. This research was initiated while S.N.F.
was visiting at the Research Center for Quantum Information,
Institute of Physics, Slovak Academy of Sciences. S.N.F. is
grateful for their very kind hospitality. This work was supported
by EU integrated project 2010-248095 (Q-ESSENCE), APVV
DO7RP-0002-10, and VEGA 2/0092/11 (TEQUDE). S.N.F. thanks the
Russian Foundation for Basic Research (projects 10-02-00312 and
11-02-00456), the Dynasty Foundation, and the Ministry of
Education and Science of the Russian Federation (projects
2.1.1/5909, $\Pi$558, 2.1759.2011, and 14.740.11.1257). M.Z.
acknowledges the support of SCIEX Fellowship 10.271 and GACR
P202/12/1142.
\end{acknowledgments}

%-------------------------------------------------------------------------------------------

\appendix
\section{Densities of states}
The densities of states $N_{E,\mu}^{\rm diff}(p)$ and
$N_{E,\mu}^{\rm same}(p)$ are introduced in
Eqs.~(\ref{density-diff}) and (\ref{density-same}), respectively.
It is worth noting that the domain of functions $N_{E,\mu}^{\rm
diff}(p)$ and $N_{E,\mu}^{\rm same}(p)$ is $P_E^-$ and $P_E^+$,
respectively. Below we present the explicit formulas of these
densities for the effects $E_{\rm asym}$, $E_{xy-}$, and $E_{z-}$
specified in Eqs.~(\ref{anti-})--(\ref{Pz}). We calculate the
densities by using either the Hilbert--Schmidt measure (\ref{HS})
or the Bures measure (\ref{bures}) and the obtained densities are
depicted in Fig.~\ref{figure:measures}.

As far as POVM effect $E_{\rm asym}$ is concerned, $P_{\rm asym}^-
= (0,\frac{1}{2}]$ and $P_{\rm asym}^+ = [0,\frac{1}{4}]$,
consequently $p_0=\frac{1}{4}$ and the density of states $N_{{\rm
asym},\mu}^{\rm diff}(p)$ is symmetrical with respect to the point
$p_0$. The density of states can be calculated explicitly in the
corresponding domains for the Hilbert--Schmidt measure and
expressed in quadratures for the Bures measure, namely,
\begin{eqnarray}
\label{density:PaHS} N_{{\rm asym},{\rm HS}}^{\rm diff}(p) &=&
\frac{9}{2} \Big[ 1+(4p-1)^2(2\ln |4p-1|-1) \Big] \, ,  \nonumber\\
\label{density:PaB} N_{{\rm asym},{\rm B}}^{\rm diff}(p) &=&
\frac{32}{\pi^2}
\int_{|4p-1|}^{1}\sqrt{\frac{r^2 - (4p-1)^2}{1 - r^2}} ~dr \, , \nonumber\\
\label{NsamePaHS} N_{{\rm asym},{\rm HS}}^{\rm same}(p) &=& 6 \sqrt{1-4p} \, , \nonumber\\
N_{{\rm asym},{\rm B}}^{\rm same}(p) &=& \frac{4 \sqrt{1-4p}}{\pi
\sqrt{p}} \, . \nonumber
\end{eqnarray}

Similarly, for the effect $E_{xy-}$ we have $P_{xy-}^- =
(0,\frac{1}{2}]$, $P_{xy-}^+ = [0,\frac{1}{4}]$, and
$p_0=\frac{1}{2}$ but the densities of states differ from those
obtained above and in the corresponding domains they read
\begin{eqnarray}
N_{{xy-},{\rm HS}}^{\rm diff}(p) &=& \frac{9}{2} \Big[
(1+2(4p-1)^2)\arccos|4p-1|\nonumber\\
&& -3|4p-1|\sqrt{1-(4p-1)^2}\Big], \nonumber\\
N_{{xy-},{\rm B}}^{\rm diff}(p) &=& \frac{16}{\pi^2}
\int_{\arcsin|4p-1|}^{\pi-\arcsin|4p-1|} d\theta \nonumber\\
&& \times \int_{\frac{|4p-1|}{\sin\theta}}^{1} \sqrt{\frac{r^2
\sin^2\theta -(4p-1)^2 }{(1 - r^2)\sin^2\theta}} ~dr,  \nonumber\\
N_{{xy-},{\rm HS}}^{\rm same}(p) &=& 12 \sqrt{p}, \nonumber\\
N_{{xy-},{\rm B}}^{\rm same}(p) &=& 4. \nonumber
\end{eqnarray}

The effect $E_{z-}$ is characterized by regions $P_{z-}^- =
(0,1]$, $P_{z-}^+ = [0,\frac{1}{2}]$, $p_0=\frac{1}{2}$ and gives
rise to the following densities of states:
\begin{eqnarray}
N_{{z-},{\rm HS}}^{\rm diff}(p) &=& \frac{9}{4}
\Big[(1+(2p-1)^2)(1-\ln |2p-1|) - 2\Big], \nonumber
\end{eqnarray}
\begin{eqnarray}
\label{density:Pz-B}N_{{z-},{\rm B}}^{\rm diff}(p) &=&
\frac{16}{\pi^2} \int_{|2p-1|}^{1} \frac{\sqrt{(r_z^2 -
(2p-1)^2)(1 - r_z^2)}}{r_z^2} ~d r_z , \nonumber\\
N_{{z-},{\rm HS}}^{\rm same}(p) &=& \frac{3p}{\sqrt{1-2p}} \, , \nonumber\\
\label{NsamePz-B}N_{{z-},{\rm B}}^{\rm same}(p) &=& \frac{4
\sqrt{2p}}{\pi \sqrt{1-2p}} \, . \nonumber
\end{eqnarray}

Also, we present the calculated densities (\ref{density-diff}) and
(\ref{density-same}) for the effect $E={\rm
diag}(\frac{1}{2},1,0,\frac{1}{2})$ of the almost universal
two-valued comparison measurement from
Sec.~\ref{section:AUC-measurement}. The result is
\begin{eqnarray}
N_{E,{\rm HS}}^{\rm diff}(p) &=& \frac{12}{5} (1-|2p-1|)^3 (|2p-1|^2+3|2p-1|+1)  \nonumber \\
N_{E,{\rm B}}^{\rm diff}(p) &=& \frac{16}{\pi^2}
 \!\! \int\limits_{-1}^{1-2|2p-1|} \!\!\!\!\!\!\!\!
\sqrt{(1-r_z^2)[1-(r_z+2|2p-1|)^2]}
~ d r_z , \nonumber \\
N_{E,{\rm HS}}^{\rm same}(p) &=& N_{E,{\rm B}}^{\rm same}(p) =
\delta\left(p-\tfrac{1}{2}\right). \nonumber
\end{eqnarray}

%-------------------------------------------------------------------------------------------

%\newpage

\end{document}